\documentclass{sig-alternate}

\begin{document}
%
\conferenceinfo{RecSys}{'14, October 6-10, 2014, Foster City, Silicon Valley, USA}

\title{The Cold-start Problem: Minimal Users' Activity Estimation}

\numberofauthors{4} 
\author{
\alignauthor
Juraj Vi\v{s}\v{n}ovsk\'{y}\\
       \affaddr{Slovak University of Technology}\\
       \affaddr{Bratislava, Slovakia}\\
       \email{visnovsky.j@gmail.com}
\alignauthor
Ondrej Ka\v{s}\v{s}\'{a}k\\
       \affaddr{Slovak University of Technology}\\
       \affaddr{Bratislava, Slovakia}\\
       \email{name.surname@stuba.sk}
\alignauthor
Michal Kompan\\
       \affaddr{Slovak University of Technology}\\
       \affaddr{Bratislava, Slovakia}\\
       \email{name.surname@stuba.sk}
\and
\alignauthor
M\'{a}ria Bielikov\'{a}\\
       \affaddr{Slovak University of Technology}\\
       \affaddr{Bratislava, Slovakia}\\
       \email{name.surname@stuba.sk}
}

\maketitle
\begin{abstract}
Cold-start problem, which arises upon the new users arrival, is one of the fundamental problems in today's recommender approaches. Moreover, in some domains as TV or multimedia - items take long time to experience by users, thus users usually do not provide rich preference information. In this paper we analyze the minimal amount of ratings needs to be done by a user over a set of items, in order to solve or reduce the cold-start problem. In our analysis we applied clustering data mining technique in order to identify minimal amount of item's ratings required from recommender system's users, in order to be assigned to a correct cluster. In this context, cluster quality is being monitored and in case of reaching certain cluster quality threshold, the recommender system could start to generate recommendations for given user, as in this point cold-start problem is considered as resolved. Our proposed approach is applicable to any domain in which user preferences are received based on explicit items rating. Our experiments are performed within the movie and jokes recommendation domain using the MovieLens and Jester dataset.
\end{abstract}



\section{Introduction}
Information overload problem is a well known and well covered by many research works. Many techniques of content filtering and personalization have been developed in order to cope with the information overload problem. Among others techniques, two stand out - collaborative filtering and content based recommendation. Both of these most popular techniques for recommendation, however, struggle when new user is introduced to the recommender system. This problem is known as the cold-start problem. The cause of the cold-start problem is lack of information about users' preferences, thus recommender system is unable to profile these users and therefore is not able to generate recommendations for the new users.

There are several techniques to reveal new user's preferences e.g. forcing user to rate certain amount of items and consequently enabling recommender system to generate relevant recommendations. Moreover, there are domains such as TV recommendation, where such a techniques cannot be applied. Similarly, we cannot define the turning point - when not enough preferences are known and the desired state, when user's preferences are known in order to generate sufficient recommendations.

The popularity of smart TV allows us to enhance standard TV services, in order to help users find relevant content over hundreds of TV channels. As a result of analysis of set-top-box logs from one of the major Slovak cable operators (Intelligent Electronic Program Guide project) we realized, that from the TV shows recommendation point of view and users experience, there is need to clearly identify cold-start problem breaking point.

In this paper we aim to find border line between cold-start and known preferences. Our goal is determination of the minimal amount of ratings performed by a user for allowing collaborative filtering algorithm to generate recommendations without being affected by cold-start.

\section{Related work}
Cold-start problem resolving in a personalized recommendation systems is currently active research topic, which resulted to the creation of a multiple different approaches to solve this problem. It is fairly conventional approach to solve cold-start problem by using hybrid recommendation techniques. In this case, it is common to combine the different recommendation algorithms for purpose of mutual elimination of the individual algorithms disadvantages. Existing types of hybrid approaches described in his work Burke~\cite{Burke}.

To tackle the cold-start problem, the most commonly used hybrid methods combine collaborative filtering and content-based filtering. An example of the successful application of this method created Leung et al.~\cite{Leung}. The authors deal with the problem of the new items occurrence. Their method primarily involves collaborative recommendation of items. If there was an item that could not be collaboratively recommended because it lacked sufficient number of user ratings, they apply to that item content-based recommendation that analyzed the relevance of the its attributes. This method reached a very good precision and recall with fewer number of unrated items, but the both metrics declined in the case where it was necessary to filter more items based on the content because of lack of sufficient number of user ratings.

A similar approach to solve the cold-start problem chose Qing and Byeong~\cite{Qing}. They use the clustering method, which is realized on the set of item attributes in system and the recommendation is based on created clusters (using a collaborative recommendation technique).

Shepitsen et al. use agglomerative hierarchical clustering to determine the similarity between users based on the tags that users assigned to movies in their system. On the basis of created clusters hierarchy, described method recommends interesting content collaboratively~\cite{Shepitsen}. Used hierarchical clustering is based on the principle that from the input set of tags is always selected the most similar pair, which is combined into a new cluster. This cluster is then reassigned to the input set from which the most similar pairs (of tags or clusters) are selected. The single clusters tree is created as the result, which is used to determine the similarity between users and then to collaborative recommendation.

The issue of new users in recommendation systems is also the objective of Zhou et al.~\cite{Zhou}. The authors tackle with this problem by training a decision tree whose nodes represented questions asked to new users. The knowledge gained by method proposed in this paper could be used to train an analogical decision tree. This kind of cold-start problem solution could be used with demographical recommendation, which is able to improve results of basic tree.

In~\cite{Kompan} the authors use group recommendation technique for single user recommendation. The user belongs to many real or even virtual groups. By using recommendation for these groups, they find recommendation for user who belongs to all used groups. The principle is used also for new user. For that user the authors create recommendation for all groups in the portal. After getting some information about the user they filter groups, which will use for him in the future.

The cold-start problem belongs to largest problems of personalized recommendation. In the appointed cases there was no evaluation of what threshold should be users taken as new and since where recommend them without restrictions. The authors set the threshold mostly by their personal judgment and they does not support it by ordinary claims.

\section{Methodology}
We proposed a methodology for discovering the minimal number of ratings that a user must perform to be assigned to the correct cluster (cluster in which the user is stable assigned so he does not migrate longer between other clusters after his next ratings).

\subsection{Data sets}
For our experiments we used data from the cinematography and entertainment domains, in order to prove that proposed methodology can be applicable to several domains with similar characteristics e.g. TV - recommended items experience take longer time, rich metadata information etc.

The first source was the set of data from the MovieLens project (Movielens 10M dataset\footnote{http://grouplens.org/datasets/movielens/}), which consists of movie (approx. 10k) ratings (approx. 10M) by multiple users (approx. 70k). For each user is given that he made at least twenty rates. 


Each movie is in this dataset characterised by unique identifier, name, release year and a set of genres which describes it. Furthermore, there is a set of users, characterized only by their unique identifier. The set of ratings consists from triples - user, item and rating. The ratings themselves constitute an explicit feedbacks received from a particular users, expressed numerically in the interval $<1, 5>$.


The second source of data was Jester Collaborative Filtering dataset 1\footnote{http://www.ieor.berkeley.edu/~goldberg/jester-data/}. This dataset consist of jokes (approx. 100) ratings (approx. 1.8M) made by approximately 25k of users. Each of them has rated between 36 and 100 items. Each user was offered by the same set of items to rate.


Dataset is composed as the rating matrix 24 983 x 102. Each row consist of user identifier, number of items he rated and one hundred item ratings. Ratings are real values ranging from - 10.0 to 10.0. Items, that user did not rate, have the rating value 99.0, which is equivalent to "not rated yet".


\subsection{Application of clustering method}
To find the optimal number of user's ratings in order to detect cold-start problem and to verify the defined assumptions (clearly determined threshold where we can stop consider user as new because we already know his preferences; increasing quality of cluster, where we affile the user after increasing the number of his ratings), we used the K-means algorithm to create cluster model of the input data based on the similar vector values of the instances.

\begin{figure}[t]
  \centering
  \includegraphics[scale=0.33]{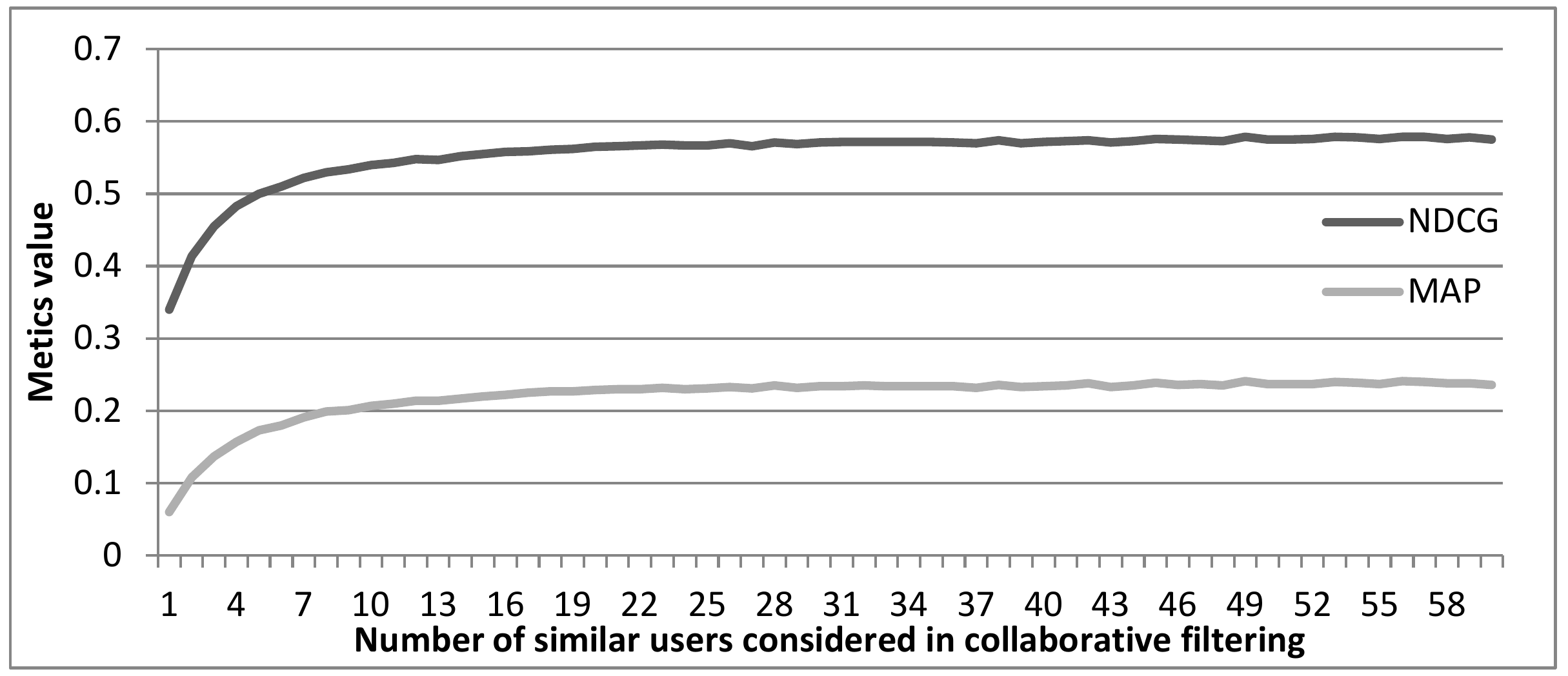}\\
  \caption{MovieLens dataset - Progress of NDCG and MAP metrics based on the number of users considered in collaborative filtering.}\label{fig1}
\end{figure}

K-means divides a set of input items on explicitly defined number of clusters. We set this number as the number of all users divided by a coefficient of $k$. Thus we achieved clusters, composed by average of $k$ users. This coefficient value $k$ was determined in respect to the results of normalized discounted cumulative gain metric (NDCG)~\cite{Ricci, Resnick} and mean average precision metric (MAP)~\cite{map} as the point at which these two metrics reached their maximum value (Fig.~\ref{fig1}). To make it clear, on the MovieLens dataset $k$ value 50, and on the Jester dataset $k$ value 100 guarantees best performing clusters (from NDCG and MAP metrics respectively).


The Bergman divergence was used in the K-means algorithm to determine average values of items in clusters and to determinate the cluster centroids. To specify the values of difference between clusters were used the Squared euclidean distance. This metric is used to the gradual optimization of process during the calculation steps. We ran each process in 10 iterations, where each of them were made up to 100 optimization steps if needed.

In the clustering process we mainly investigate the quality of individual clusters. The quality of clusters determines in this case the similarity of cluster inside items and dissimilarity of the outer items. For this purpose we used the Distance cluster performance metric, which is based on the Davies Bouldin index~\cite{dav-bou}.

The Davies Bouldin index metric~(Eq.~1) determines the average distance from cluster items to the cluster centroid. The output of this metric is a negative decimal number that, with increasing cluster quality, converges to the value of 0.

\begin{equation}\small
DB(C)=\frac{1}{c}
\sum_{j=1}^{k}max(\frac{\Delta{C_{j}}+\Delta{C_{m}}}{\delta(C_{j},c_{m})})\: m!=j,
\end{equation}

where $C_{i}$ represents the $i^{th}$ cluster, $\Delta C_{i}$ variance in the$i^{th}$ cluster and $\delta$ Euclidean distance between pair of clusters.

We managed to apply method to randomly chosen subset of 100 users because of the high computation cost. Each of chosen users rated at least 50 ratings, thus we can guarantee the same information value for each evaluation step.

\subsection{Evaluation}
In this work we focused on exploration of the dependency between the clusters quality and the number of user ratings. Next, we have analyzed the success rate of finding the final cluster for each user after certain amount of his ratings.

The first aspect we examine is the monitoring of the minimum number of ratings that must a user perform to ensure, that we are able to assign him to the correct cluster. Correctness of the cluster assignment has been determined through comparison to the final cluster (cluster which is assigned to the user when all his rating history is considered). Our comparisons were realized incrementally, with the increasing number of user ratings included (both datasets).

\begin{figure}[t]
  \centering
  \includegraphics[scale=0.44]{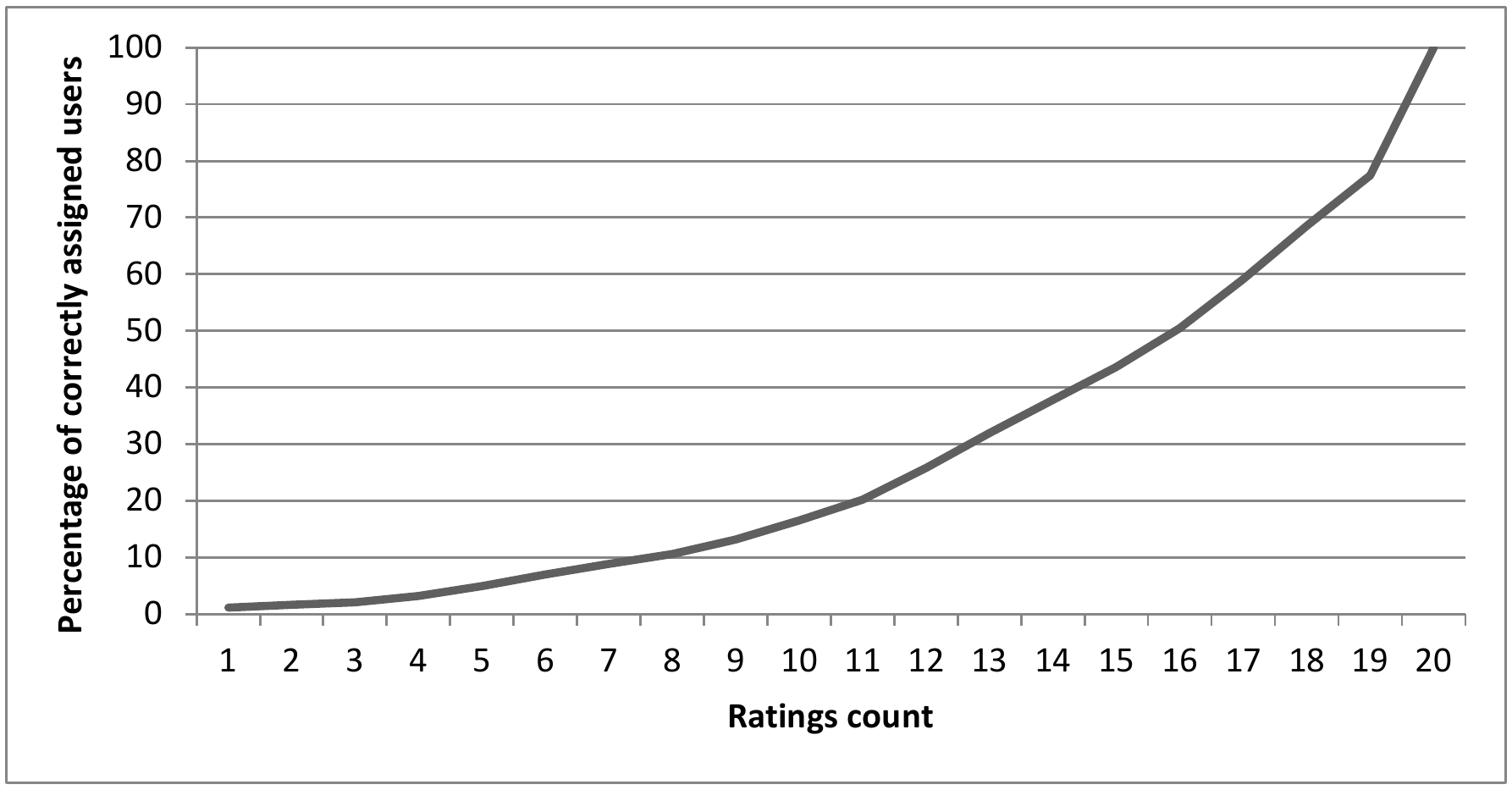}\\
  \caption{MovieLens dataset: Percentage of users associated with correct cluster after first twenty ratings.}\label{fig2}
\end{figure}

\begin{figure}[t]
  \centering
  \includegraphics[scale=0.44]{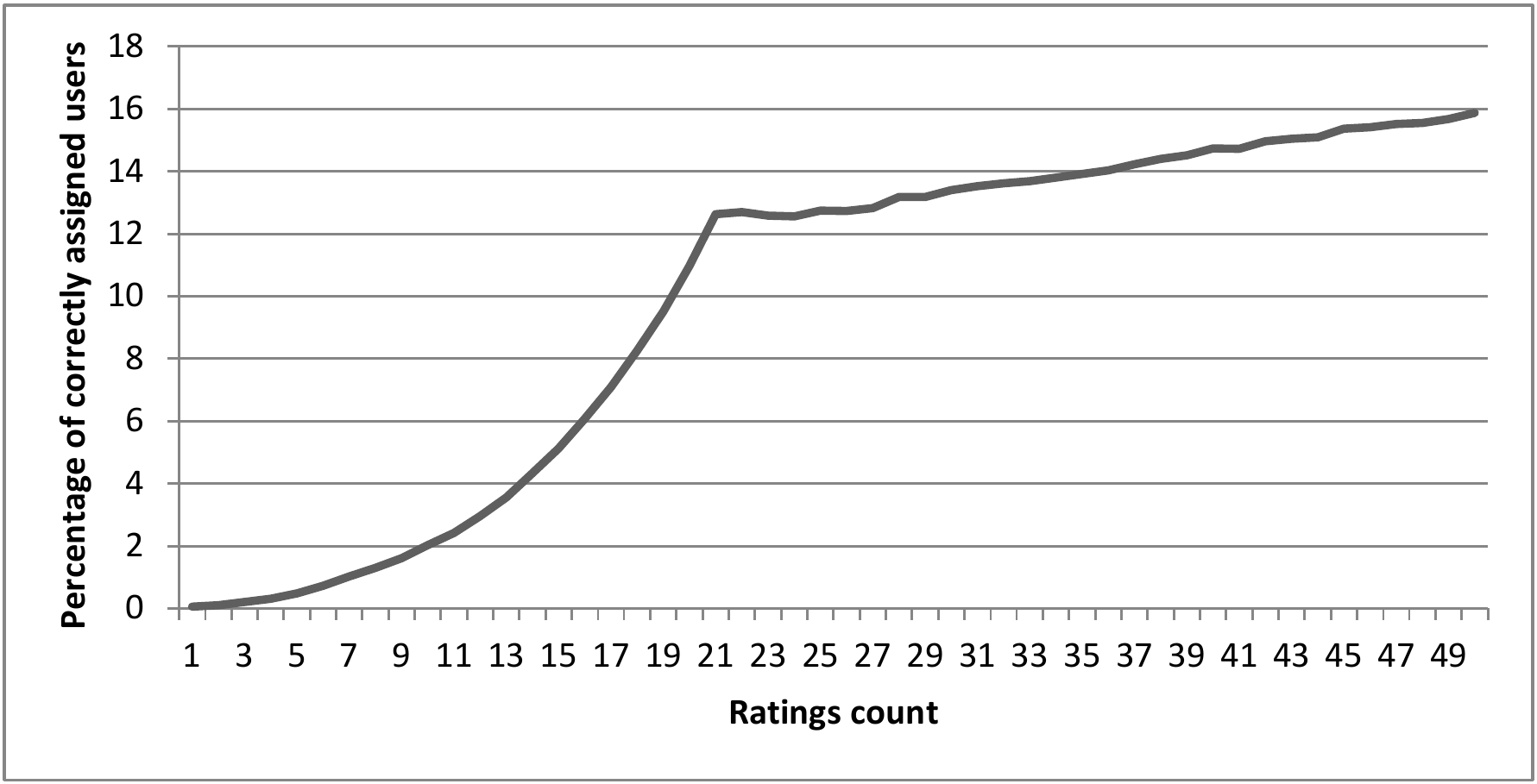}\\
  \caption{MovieLens dataset: Percentage of users associated with correct cluster after certain amount of ratings (up to 50).}\label{fig3}
\end{figure}

\begin{figure}[t]
  \centering
  \includegraphics[scale=0.44]{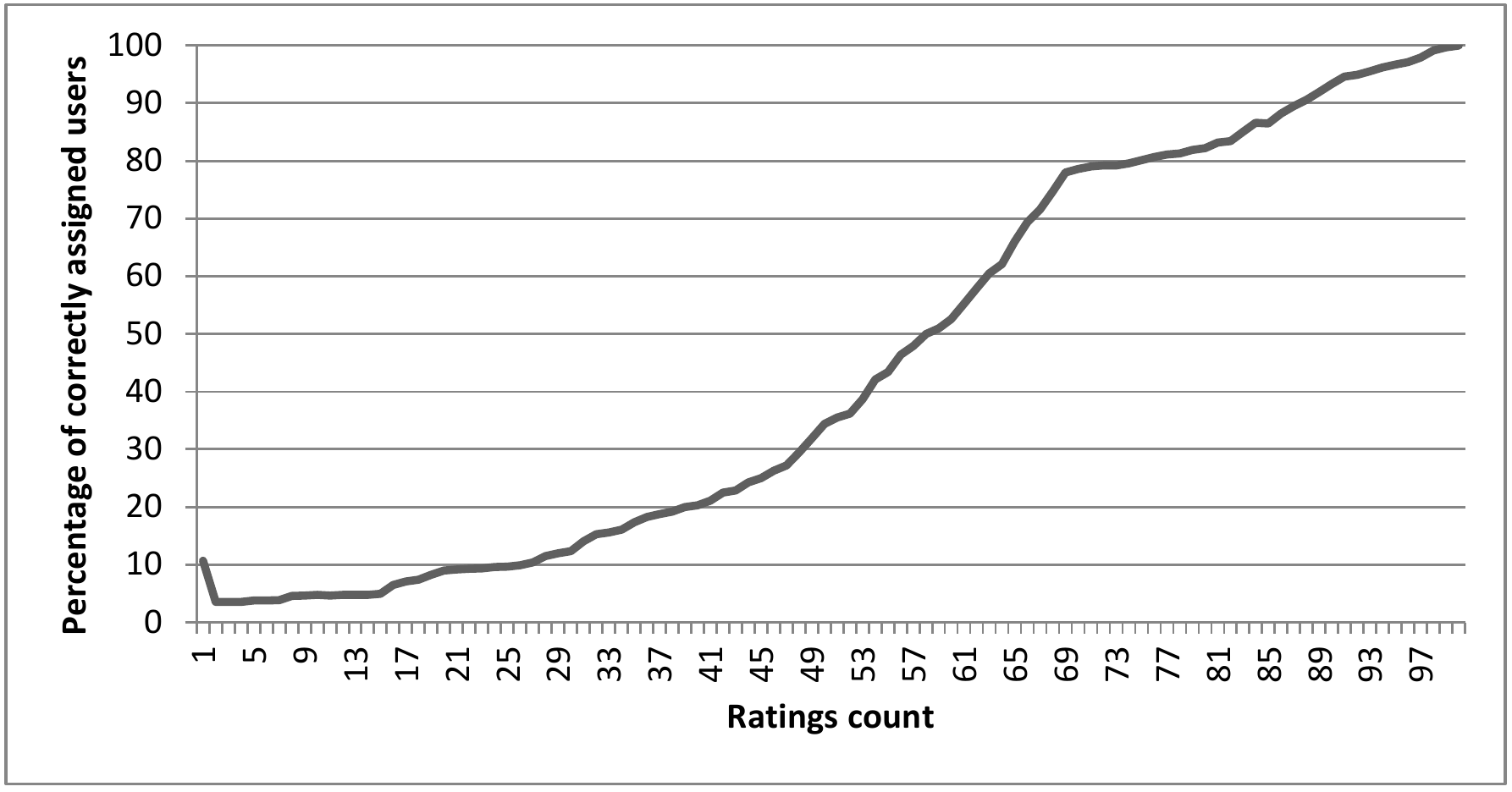}\\
  \caption{Jester dataset: Percentage of users associated with correct cluster after certain amount of ratings (up to 100).}\label{fig4}
\end{figure}

The results of MovieLens dataset are shown in Fig.~\ref{fig2} and Fig.~\ref{fig3}, the results of Jester dataset are shown in Fig.~\ref{fig4}. Experiment over MovieLens dataset was divided into two parts. Reason for this split was, that 2.7\% (1 902) from all users (69 878) made exactly the minimal prescribed number of ratings (20), which created unbalanced peak at this point. Other users' rating counts, however, occurred in significantly lower counts distributed from 21 ratings to several hundreds. Users with minimal ratings count would influence results for first 20 ratings if were included.

The results achieved for users with exactly 20 ratings are therefore shown individually in Fig.~\ref{fig2}. During the whole experiment, we can observe an exponential growth of cluster assignment success. This situation represents an ideal situation, which in fact does not occur in real, proved by experiments with remaining MovieLens users and Jester dataset.

Trend of successful cluster assignments for the remaining MovieLens users (the ones with more than 20 ratings) is shown in Fig.~\ref{fig3}. As we can see, at the beginning (only few ratings considered) the percentage of successful association grows exponentially, as in previous case. This trend occurs only until a certain threshold (21 ratings), where it changes into a linear increase.  Similar trend of association success was observed within Jester dataset (Fig.~\ref{fig4}). At the beginning, there can be seen an exponential success growth, which, however, at some point changes into the linear increase. In this case is the minimum number of ratings, that users had to perform, set at the level of 36, but the change of association success growth occurs after approx. 68 ratings.

Despite the different absolute values, we can say that the association success growth has the same tendency for both tested datasets. From this, we can deduce some interesting conclusions. For the both datasets we have in fact found, that up to a certain point increase the added value of each new item rating made by user exponentially. To this point is therefore preferred to require from the user further and further ratings, since he can obtain a rich preference data. After this point does not the amount of preference data constitute such a significant contribution and therefore we should consider whether the negatives arising from user bothering do not exceed obtained added value from preference data. Here we refer to overcome the cold-start phase, when we require from the user a package of initial information, while we do not give him back any added value from recommendation yet. Feedback form the user actions is however collected all the time, regardless of the intensity value added growth.

\begin{figure}[t]
  \centering
  \includegraphics[scale=0.44]{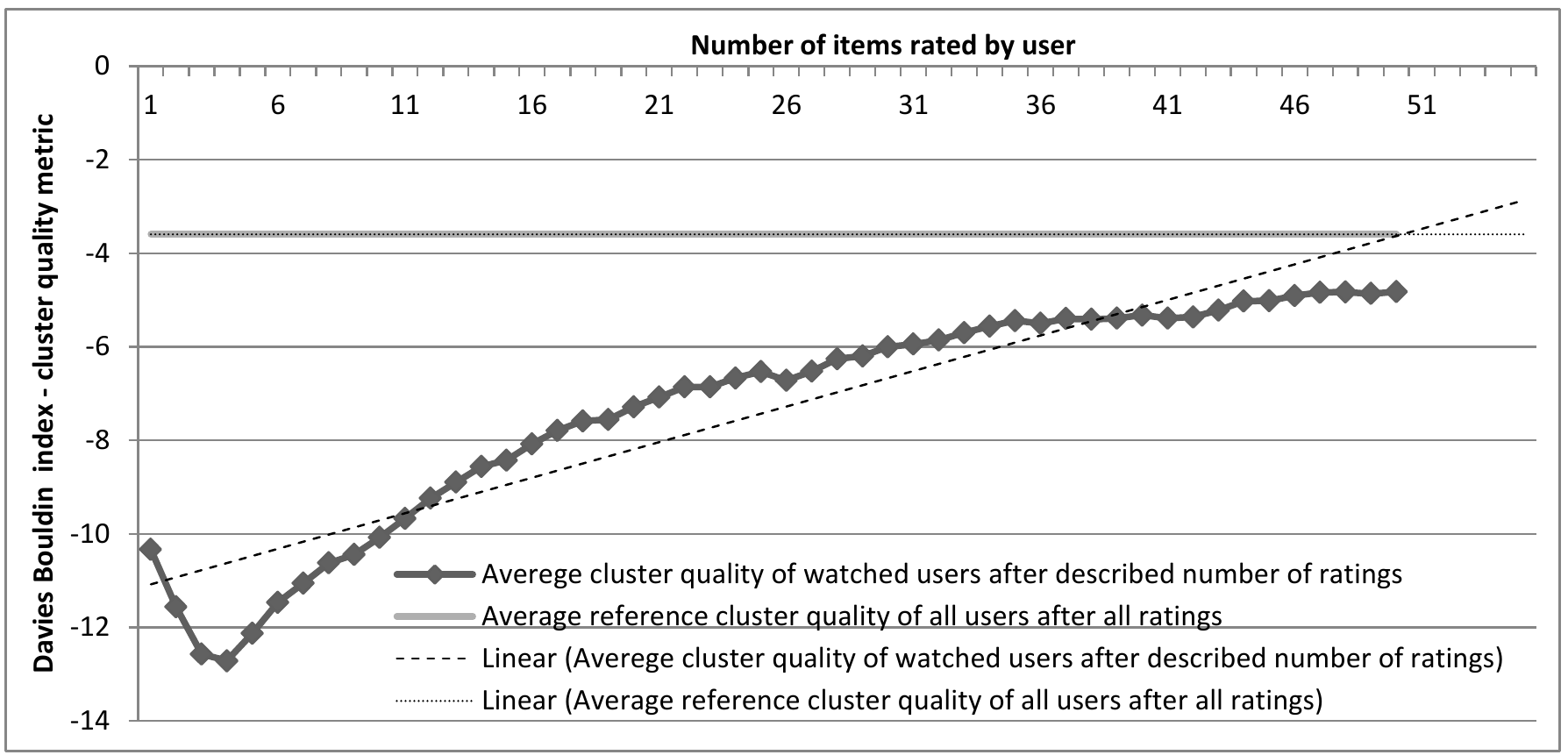}\\
  \caption{Dependence between cluster quality and the number of ratings considered per user.}\label{fig5}
\end{figure}

Next, we focused on the average quality of clusters, in which we assign users after a certain number of their ratings (Fig.~\ref{fig5} - MovieLens dataset, similar pattern was observed for Jester dataset). This experiment for reveals several facts. First, it shows that with increasing number of user ratings increases in average also the quality of a cluster in which the user is actually assigned. Its value increases logarithmically and converges to the reference clusters average value. As a reference we mark clusters in which the users are assigned after count all of their ratings.

The moment when the current cluster quality reaches the average value of the reference clusters is not possible to determine precisely, because the actual cluster's quality value and reference value converge together. After 50 ratings is the quality value of the user's actual cluster at level -4.81734 of Davies Bouldin index and the average clusters quality is at level -3.58857 of Davies Bouldin index. But from Fig.~\ref{fig5} we can see that the regression curves of both curves are intersected at the level of 50 ratings.

\section{Conclusions}
In this work, we aimed to determine the influence of new user ratings to overcome the cold-start problem. We have worked with an explicit feedback in the form of users' item ratings. Base on ratings, users have been clustered. We explored how the clustering process for user stabilizes with a growing number of his ratings available. Our experiments were performed on two datasets from different domains, in order to investigate domain dependency of obtained results and recommendations. Due to the fact that we were able to demonstrate the same behavior on two independent datasets from different domains, we can generalize our findings. Such information we use to enhance TV show recommendation, as we can use different weights of various recommendations approaches (weighted hybrid approach).

Experiments revealed that the preference data collected as the user feedback has for both datasets a similar behavior. Despite the different absolute values reached for individual datasets, we can conclude that the association success growth has the same tendency. From this, we deduced some interesting conclusions. For the both datasets we have found that up to a certain point increases added value of each new user's item rating exponentially. To this point is therefore preferred to require from user further and further ratings, since they help to obtain a rich preference data.

After this point, the growth slows down to the linear increase. A point where the added value from obtained preference data growth changes is individual and must be determined by described methodology for every domain. We believe, that in the phase of overcoming the cold-start, it is suitable to collect feedback only to described point. Consequently, it is appropriate to begin recommend to user in this phase and this way reward him for his previous feedback. Nevertheless we continue to receive feedback from user, although it gives us much little valuable information as for the new user.

\section{Acknowledgements}
This work was partially supported by the Scientific Grant Agency of the Slovak Republic, grant No. VG1\//0675\//11 and the Slovak Research and Development Agency under the contract No. APVV-0208-10.

\bibliographystyle{abbrv}

\end{document}